\documentclass[journal,twoside,web]{ieeecolor}
\usepackage{generic}
\usepackage{graphicx}
\usepackage{epstopdf}
\usepackage[caption=false]{subfig}

\usepackage{amsmath,amssymb,amsfonts,amsthm}

\usepackage[ruled,vlined]{algorithm2e}

\DeclareMathOperator*{\argmin}{arg\,min}

\DeclareMathOperator*{\diag}{diag}

\newtheorem{theorem}{Theorem}
\newtheorem{corollary}{Corollary}

\newtheorem{remark}{Remark}

\newcommand{\squeezeup}{\vspace{-2.5mm}}

\begin{document}
\title{A Contraction-constrained Model Predictive Control for Multi-timescale Nonlinear Processes \thanks{This work was partially supported by Australian Research Council Discovery Project DP210101978.}}

\author{Ryan~McCloy,~Lai Wei,~Jie~Bao
\thanks{R. McCloy, L. Wei, and J. Bao are with the School of Chemical Engineering, The University of New South Wales, Sydney, NSW 2052, Australia. E-mail: j.bao@unsw.edu.au (corresponding author).}}
\maketitle
\thispagestyle{empty}
\begin{abstract}
Many chemical processes exhibit diverse timescale dynamics with a strong coupling between timescale sensitive variables. Model predictive control with a non-uniformly spaced optimisation horizon is an effective approach to multi-timescale control and offers opportunities for reduced computational complexity. In such an approach the fast, moderate and slow dynamics can be included in the optimisation problem by implementing smaller time intervals earlier in the prediction horizon and increasingly larger intervals towards the end of the prediction. In this paper, a reference-flexible condition is developed based on the contraction theory to provide a stability guarantee for a nonlinear system under non-uniform prediction horizons.
\end{abstract}
\begin{keywords}
   Multiple timescales, nonlinear model predictive control, discrete-time nonlinear systems, contraction theory.
\end{keywords}

%%%%%%%%%%%%%%%%%%%%%%%%%%%%%%%%%%%%%%%%%%%%%%%%%%%%%%%%%%%%%%%%%%%%%%%%%%%%%%%%%%%%%%%%%%%%%%%%%%%%%%%%%%%%
\section{Introduction}
Model Predictive Control (MPC) has become the industry standard for model-based optimal control, with widespread applications and technical developments \cite{qin2003survey}. The objective of an MPC is to solve a finite horizon optimisation problem with respect to some constraints \cite{fontes2018guaranteed,RaM09}. In chemical industries, MPC has seen great success due to its ability to deliver ``optimal'' performance and to explicitly handle constraints. Often, a trade-off is made between prediction horizon length and the use of detailed models (attracting increased computational burdens) against the acceptable performance of the controller. 

Whilst many chemical processes can be modelled with a high level of accuracy by using more than one time scale \cite{chang1984multi,bailey2018biochemical}, the use of such models can be computationally challenging \cite{pinnamaraju2018empirical}. In some cases, this introduced complexity is necessary since controller design using a model in only one timescale may cause instability when applied to the physical system. It is usually also true for such systems that a long prediction horizon (e.g., in the case of MPC) is needed for effective control. However, due to computational limitations, long optimisation horizons combined with relatively short time intervals are generally impractical \cite{findeisen2016exploiting}.

To overcome these complexity issues, one approach is to modify the standard MPC formulation for the multi-timescale setting \cite{tan2016model}, by introducing timescale specific models (using non-uniform time intervals) and hence prediction horizons. In this approach, small time intervals (capturing fast process dynamics) and large time intervals (capturing slower dynamics) are used to ensure short term accuracy whilst facilitating long term predictions. As a result, the computational burden associated with detailed predictions over long horizons can be significantly reduced by using models with larger time intervals without sacrificing significant accuracy or introducing significant complexities. Traditional methods to ensure the closed-loop stability of MPC schemes involve using the cost as a Lyapunov function \cite{RaM09}, which relies on models with uniform time intervals, and hence a uniformly spaced optimisation horizon to show that the cost is decreasing.

If arbitrary cost functions are considered, this stability issue is further compounded and consequently requires a tailored stability condition \cite{ellis2017economic}. In addition, if the reference trajectory is considered time-varying, stability guarantees for an optimal controller require a condition that is also reference-independent, e.g., those based on incremental stability \cite{angeli2002lyapunov}. Introduced by \cite{lohmiller1998contraction}, contraction theory facilitates stability analysis and control of nonlinear systems with respect to time-varying (feasible) references without redesigning the control law \cite{manchester2017control}. Contraction theory is particularly attractive in that it can be used to analyse the incremental stability of nonlinear systems and simultaneously synthesise a tracking controller using control contraction metrics (see, e.g., \cite{manchester2017control}). This has motivated developments for contraction-based MPCs \cite{mccloy2021differential}, offering significant flexibility over Lyapunov-based alternatives. 

Inspired by the work in \cite{tan2016model}, we propose a non-uniformly spaced optimisation horizon nonlinear MPC with timescale specific contraction-based stability constraints. Extending from \cite{tan2016model}, we relax the linear system requirements to consider discrete-time nonlinear systems and propose a novel \textit{modular} contraction-based stability constraint to ensure closed-loop stability across multiple time intervals (for multiple timescale specific models). Moreover, this construction is reference-flexible and scalable, permitting tracking of \textit{a priori} unknown time-varying operation targets, arbitrary economic cost functions, and inclusion of additional timescale models in the MPC without increasing the computational burden.     

The remainder of this article is structured as follows, Section \ref{sec:pre} presents the prerequisite contraction theory tools, and Section \ref{sec:pro} formulates the multi-timescale MPC problem. Section \ref{sec:ctr} develops contraction-based controller structures for systems subject to multiple timescales and imposes them as constraints on a multi-timescale MPC. Section \ref{sec:exa} demonstrates the overall method via numerical simulation and Section \ref{sec:conclusion} concludes this article. 

\section{Preliminaries} \label{sec:pre}
\begin{figure} \label{fig:rie_geo}
    \begin{center}
        \includegraphics[width=\linewidth]{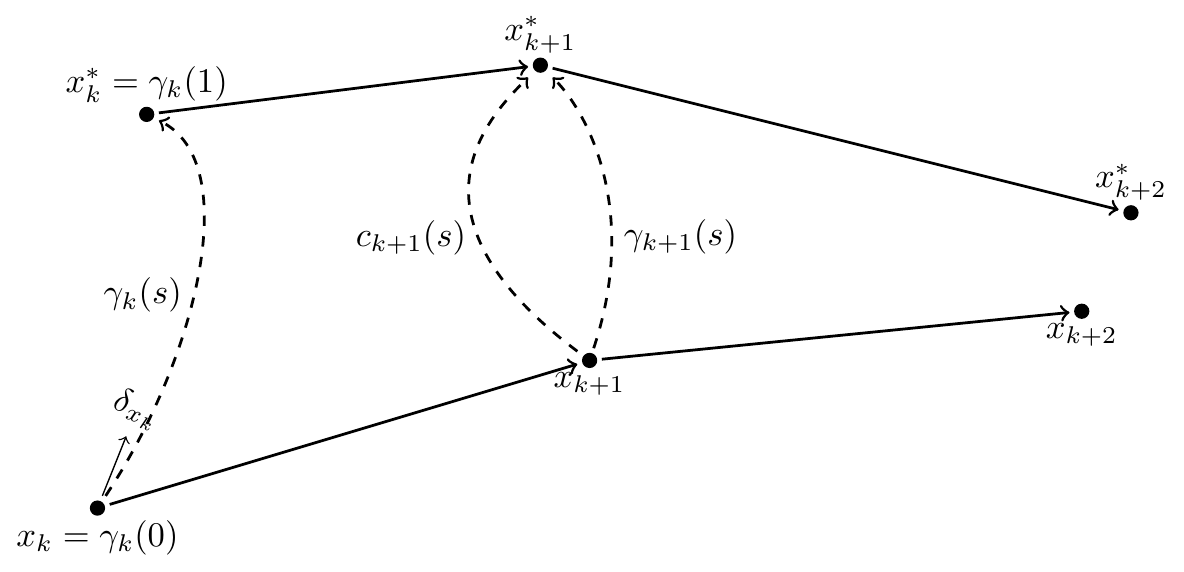}
        \caption{State and reference trajectories with $s$-parameterised paths.}
    \end{center}
    \squeezeup\squeezeup
\end{figure}

Firstly, consider the single timescale nonlinear system
\begin{equation}\label{equ:pre cer sys}
    x_{k+1} = f(x_k) + g(x_k)u_k,
\end{equation}
where state and control are $x_k \in \mathcal{X} \subseteq \mathbb{R}^n$ and $u_k \in \mathcal{U} \subseteq \mathbb{R}^m$. The corresponding differential system of \eqref{equ:pre cer sys} is
\begin{equation}\label{equ:pre cer dif sys}
    \delta_{x_{k+1}} = A\delta_{x_k} + B\delta_{u_k},
\end{equation}
where Jacobian matrices of $f$ and $g$ in \eqref{equ:pre cer sys} are defined as $A:=\frac{\partial (f(x_k) + g(x_k)u_k)}{\partial x_k}$ and $B:=\frac{\partial (f(x_k) + g(x_k)u_k)}{\partial u_k}$  respectively, $\delta_{u_k} := \frac{\partial u_k}{\partial s}$ and $\delta_{x_k}:=\frac{\partial x_k}{\partial s}$ are vectors in the tangent space $T_x\mathcal{U}$ at $u_k$ and tangent space $T_x\mathcal{X}$ at $x_k$ respectively, where $s$ parameterises a path, $c(s): [0,1] \rightarrow \mathcal{X}$ between two points such that $c(0) = x, c(1) = x^*  \in \mathcal{X}$ (see Fig. \ref{fig:rie_geo}). Consider a feedback control law for the differential dynamics \eqref{equ:pre cer dif sys} as
\begin{equation}\label{equ:pre dif fed}
    \delta_{u_k} = K(x_k) \delta_{x_k},
\end{equation}
where $K$ is a state dependent feedback gain. The contraction condition for a discrete-time nonlinear control affine system in \eqref{equ:pre cer sys} can be described as follows,
\begin{theorem}[\cite{wei2021contractionsyntehsis}]\label{thm:pre ctr con}
    For a discrete-time nonlinear system \eqref{equ:pre cer sys}, with differential dynamics \eqref{equ:pre cer dif sys} and differential state-feedback controller \eqref{equ:pre dif fed}, provided a uniformly metric, $\underline{m}I \leq M(x_k) \leq \overline{m}I$, exists satisfying, 
    \begin{equation}\label{equ:pre ctr con} 
        (A+BK)^\top M_+(A+BK) - (1-\beta)M < 0,
    \end{equation}
    where $M_+=M(x_{k+1})$, the closed-loop system is contracting for some constant $0 < \beta \leq 1$. Furthermore, the closed-loop system is exponentially incrementally stable,  i.e.,
    \begin{equation}
        \label{equ:pre exp sta}
        |x_k-x^*_k| \leq  R e^{-\lambda k} |x_0 - x^*_0|,
    \end{equation}    
    for some constant $R$, convergence rate $\lambda$ and any feasible reference trajectory $(x^*, u^*)$, satisfying \eqref{equ:pre cer sys}, where $x_0$ is the initial state and $x_k$ is the state at time-step $k$.

    % \begin{enumerate}
    %     \item 
    % \begin{equation}
    %     \label{equ:pre asy sta}
    %         |x_k-x^*_k| \leq (1-\beta)^{\frac{k}{2}} R|x_0 - x^*_0|,
    %     \end{equation}
    %     for some constant $R$, where $x_k$ is the $k$-th step of the closed-loop state trajectory, and
    %     \item incrementally exponentially stable, i.e.,
    %     \begin{equation}
    %     \label{equ:pre exp sta}
    %         |x_k-x^*_k| \leq  R e^{-\lambda k} |x_0 - x^*_0|,
    % \end{equation}
    % for some constants $R$ and $\lambda$.
    % \end{enumerate}
\end{theorem}
If condition \eqref{equ:pre ctr con} holds for all feasible state values in a region, that region is called a contraction region. In Theorem \ref{thm:pre ctr con}, the metric, $M$, is used in describing Riemannian geometry, which we briefly introduce here. We define the Riemannian distance, $d(x,x^*)$, as (see, e.g., \cite{do1992riemannian})
\begin{equation}\label{equ:Riemannian distance and energy}
    \begin{aligned}
    d(x,x^*) = d(c) :=\int_0^1 \sqrt{\delta^\top_{c(s)}M(c(s))\delta_{c(s)}}ds,
    \end{aligned}
\end{equation}
where $\delta_{c(s)} := \frac{\partial c(s)}{\partial s}$.
The shortest path in Riemannian space, or \textit{geodesic}, between $x$ and $x^*$ is defined as 
\begin{equation}\label{equ:geodesic}
    \gamma(s) :=\argmin_{c(s)} {d(x,x^*)}.
\end{equation}

Leveraging Riemannian tools, a feasible contracting tracking controller controller for \eqref{equ:pre cer sys}, is obtained by integrating the differential feedback law \eqref{equ:pre dif fed} along the geodesic, $\gamma(s)$ \eqref{equ:geodesic}, as
\begin{equation}\label{equ:pre ctl int}
    u_k = u^*_k + \int_0^1K(\gamma(s))\frac{\partial \gamma(s)}{\partial s}\,ds.
\end{equation}

In summary, a contracting tracking controller controller ensures that the length of the minimum path (i.e., geodesic) between any two trajectories (e.g., the plant state, $x$, and desired state, $x^*$, trajectories), with respect to the metric $M$, shrinks with time, i.e., provided that the contraction condition \eqref{equ:pre ctr con} holds for the discrete-time nonlinear system \eqref{equ:pre cer sys}, %\LAI{in the contraction region}, 
we can employ a stabilizing feedback controller \eqref{equ:pre ctl int} to ensure convergence to feasible operating targets.

%%%%%%%%%%%%%%%%%%%%%%%%%%%%%%%%%%%%%%%%%%%%%%%%%%%%%%%%%%%%%%%%%%%%%%%%%%%%%%%%%%%%%%%%%%%%%%%%%%%%%%%%%%%%
\section{Problem Formulation} \label{sec:pro}

\subsection{Multiple Timescale Nonlinear Processes}
In many chemical processes, the timescales of the underlying dynamics are distributed over a very large range. As such, the process dynamics can be described as functions of appropriate time intervals. Herein, we consider the timescale dependent nonlinear process model 
\begin{equation}\label{eq:sys_dist_nom}
    x_{k+1} = f(x_k,\tau) + B(\tau) u_k 
\end{equation}
where $\tau$ denotes the discrete time interval. For processes described by \eqref{eq:sys_dist_nom}, the process dynamics vary based on the time interval in which it is viewed, e.g., fast dynamics over short intervals and slow dynamics over long time intervals. The primary goal is to drive \eqref{eq:sys_dist_nom} to feasible target trajectories $x^*_k$, described as \begin{equation}\label{eq:sys_ref_nom}
    x^*_{k+1} = f(x^*_k,\tau) + B(\tau) u^*_k,
\end{equation}
where the sequence $(x^*_{k+1},x^*_k,u^*_k)$ is a solution to \eqref{eq:sys_dist_nom} (note that in the following sections, employing an MPC negates the need to solve for $u^*_k$).

\subsection{MPC using Multiple Timescale Models}
For processes described by \eqref{eq:sys_dist_nom} to ``optimally'' track time-varying operating targets, i.e. $x_k\rightarrow x_k^*$, the incorporation of a prediction based control strategy, such as MPC, is naturally befitting. In order to design an MPC, the optimisation horizon must be long enough to cover the dominant slow dynamics (with long time constant), with a time interval small enough to capture the fast dynamics. In general, this is relatively impractical due to computational limitations. One approach is to incorporate multiple timescale models, which allows the optimisation horizon to capture the slower process dynamics without significantly increasing computational complexity. The controller, $u_k$, can then be computed to additionally optimise system economy via the minimisation of an arbitrary cost function, $\ell$, i.e., an MPC solves the following multi-timescale (multi-model) optimisation problem,
\begin{equation}
\begin{aligned}\label{eq:EMPCini}
\quad \min_{\hat{u}} & \sum_{i=0}^{N} \ell (\hat{x}_i,\hat{u}_i,\tau_i), \\
\text{s.t.} \quad & \hat{x}_0 = x_k,  \\
\quad & \hat{x}_{i+1} = f(\hat{x}_i,\tau_i) + B(\tau_i)\hat{u}_i, \\
& \hat{u}_i \in \mathcal{U}, \quad \hat{x}_i \in \mathcal{X}, %\\ %&r(\hat{x}_i,x^*_i,\hat{u}_i,\tau_i) \leq 0
\end{aligned}
\end{equation}
where $N$ is the prediction horizon, $\hat{x}_i$, and $\hat{u}_i$ are the respective $i$-th step state and control predictions, computed using the time interval $\tau_i$ and corresponding modelled functions $f(\tau_i)$ and $B(\tau_i)$. $x_k$ is the current state measurement, $\mathcal{U},\mathcal{X}$ are the sets of permissible inputs and states respectively.

The time interval, $\tau_i \in \mathbb{R}^+$, is varied after $N_i$ steps during the optimisation horizon according to the process dynamics, to account for the multiple timescale nature of the process. This causes the discrete-time optimisation model of the process to vary (non-uniformly) with the time interval. For brevity, we consider $p$ possible time intervals and note that the time intervals $\tau_i$ in the optimisation horizon are unrelated to the physical sampling period (e.g., denoted by $\tau_\Delta$) between which process measurements are sampled and control actions are applied, which remains constant (i.e., equal to $\tau_\Delta$).

Assuming feasibility of the optimisation problem~\eqref{eq:EMPCini}, the optimal input trajectory, $\mathbf{\hat{u}}^{opt} = (\hat{u}^{opt}_{0}, \hat{u}^{opt}_{1}, \cdots, \hat{u}^{opt}_{N-1}) \in \mathcal{U}^{N}$, can be computed. The MPC is then implemented in receding horizon fashion, by applying the first control action $\hat{u}^{opt}_{0}$ to system \eqref{eq:sys_dist_nom} until the next step, $k+1$, i.e.,
\begin{equation}
\label{eq:EMPCu}
u(k) = \hat{u}^{opt}_{0} : [k,k+1).
\end{equation}

There are two key obstacles in finding solutions to the optimisation problem in \eqref{eq:EMPCini}:
\begin{enumerate}
    \item[i.] Assessing stability over non-uniform time intervals.
    \item[ii.] Ensuring convergence of solutions to time-varying reference trajectories under arbitrary cost functions.
\end{enumerate}
Traditional methods to ensure the stability of MPC schemes involve using the cost function as a Lyapunov function \cite{RaM09}. This approach relies on models with uniform time intervals, and hence a uniformly spaced optimisation horizon to show that the cost is decreasing. This problem is further compounded when arbitrary cost functions are considered, whereby the cost function can be utilised to represent the process economy. Optimal solutions can therefore be associated with state values that deviate from target operating points. Since arbitrary cost functions are considered, a stability condition is required \cite{ellis2017economic}. As the reference trajectory is considered time-varying, stability guarantees for an optimal controller require a condition that is also reference-independent (incremental stability).
As a consequence, the MPC \eqref{eq:EMPCini} requires a contraction based constraint or equivalent, as will be demonstrated in the following sections.

\subsection{Objective and Approach}
To achieve the objective of state trajectory tracking, i.e., $x_k \to x^*_k$, whilst optimising system economy, for multi-timescale nonlinear processes, we require an MPC with conditions that can ensure convergency along the full prediction horizon under each timescale. To this effect, we propose to:
\begin{enumerate}
    \item[i.] Determine timescale dependent tractable stabilising controller conditions using contraction metrics
    \item[ii.] Impose stability conditions as constraints on an MPC
\end{enumerate}
In this way, tractable offline stability conditions can be constructed and imposed online on an MPC, which has the capacity for incorporating multiple timescale models whilst simultaneously optimising an economic cost and ensuring tracking of operating targets.  

%%%%%%%%%%%%%%%%%%%%%%%%%%%%%%%%%%%%%%%%%%%%%%%%%%%%%%%%%%%%%%%%%%%%%%%%%%%%%%%%%%%%%%%%%%%%%%%%%%%%%%%%%%%%
\section{Contraction-constrained MPC using Multiple Timescales}\label{sec:ctr}
In the following section, the contraction theory framework is leveraged to derive stability conditions to ensure closed-loop objective tracking of \eqref{eq:sys_dist_nom} over specific timescales. These conditions are transformed into tractable synthesis conditions and eventually used to characterise the property of stabilising controllers, forming the necessary MPC stability constraints.

%%%%%%%%%%%%%%%%%%%%%%%%%%%%%%%%%%%%%%%%%%%%%%%%%%%%%%%%%%%%%%%%%%%%%%%%%%%%%%%%%%%%%%%%%%%%%%%%%%%%%%%%%%%%
\subsection{Timescale Dependent Control Contraction Metrics for Multiple Timescale Nonlinear Systems}
Since varying the time interval, $\tau$, results in multiple, timescale specific models of the process, we rewrite the dynamics in \eqref{eq:sys_dist_nom} using the following convenient state model
\begin{equation}\label{eq:sys_nom_taus}
    x_{\tau,k+1} = f_\tau(x_{\tau,k}) + B_\tau u_{\tau,k},
\end{equation}
and analogously for the reference model
\begin{equation}\label{eq:sys_ref_taus}
    x^*_{\tau,k+1} = f_\tau(x^*_{\tau,k}) + B_\tau u^*_{\tau,k}.
\end{equation}
Naturally, the corresponding differential dynamics for \eqref{eq:sys_nom_taus} are also timescale specific, described by
\begin{equation}\label{eq:dsys_nom_taus}
    \delta_{x_{\tau,k+1}} = A_\tau\delta_{x_{\tau,k}} + B_\tau\delta_{u_{\tau,k}},
\end{equation}
where $A_\tau := \frac{\partial f_\tau}{\partial x}$, and we consider a suitable differential feedback controller
\begin{equation}\label{eq:dsys_fed_taus}
    \delta_{u_{\tau,k}} = K_\tau(x_{\tau,k}) \delta_{x_{\tau,k}}.
\end{equation}
We then have the following direct result from Theorem \ref{thm:pre ctr con}.
\begin{corollary}\label{cor:cc_taus}
For each time interval, $\tau$, given the discrete-time nonlinear system \eqref{eq:sys_nom_taus}, with differential dynamics \eqref{eq:dsys_nom_taus} and differential state-feedback controller \eqref{eq:dsys_fed_taus}, provided uniformly bounded metrics, $\underline{m}_\tau I \leq M_\tau(x_{\tau,k}) \leq \overline{m}_\tau I$, exist satisfying, 
    \begin{equation}\label{eq:dist ccon} 
        (A_\tau+B_\tau K_\tau)^\top M_{\tau,+}(A_\tau+B_\tau K\tau) - (1-\beta_\tau)M_\tau < 0,
    \end{equation}
    where $M_{\tau,+}=M_\tau(x_{\tau,k+1})$, the differential system is contracting with rate $0 < \beta_\tau \leq 1$ on discrete intervals of $\tau$. Furthermore, over such intervals, the nonlinear system is exponentially incrementally stabilisable.
\end{corollary}
\begin{proof}
Suppose there exists a smooth feedback controller, e.g., with general timescale dependent structure $u_{\tau,k} = k_\tau(x_{\tau,k})+\bar{u}_{\tau,k}$, with a corresponding differential form as in \eqref{eq:dsys_fed_taus}, i.e., $\frac{\partial k_\tau}{\partial x} = K_\tau$. Then, substituting \eqref{eq:dsys_fed_taus} into \eqref{eq:dsys_nom_taus} yields, the closed-loop differential system for \eqref{eq:sys_nom_taus} as
    \begin{equation}\label{eq:diffcl_d}
        \delta_{x_{\tau,k+1}} = \left(A_\tau(x_{\tau,k}) + B_\tau K_\tau(x_{\tau,k})\right)\delta_{x_{\tau,k}}.
    \end{equation}
Substituting the closed-loop differential dynamics \eqref{eq:diffcl_d} and timescale specific differential Lyapunov function,
\begin{equation}\label{equ:generalised distance}
    V_\tau(x_{\tau,k},\delta_{x_{\tau,k}}) = \delta_{x_{\tau,k}}^\top M_\tau(x_{\tau,k})\delta_{x_{\tau,k}},
\end{equation}
into the exponential discrete-time Lyapunov condition of 
\begin{equation}
    \label{eqe:Vdotcondition}
    V_{\tau,k+1} - (1-\beta_\tau)V_{\tau,k} < 0,
    \end{equation}
provides the timescale specific contraction condition in \eqref{eq:dist ccon}. To see that satisfaction of this condition implies stabilisability over each time interval, note that \eqref{eq:dist ccon} is a timescale specific instance of \eqref{equ:pre ctr con} in Theorem \ref{thm:pre ctr con}.
\end{proof}
The condition in \eqref{eq:dist ccon} can then be transformed into a tractable synthesis condition by adapting the results of \cite{wei2021control}.
\begin{corollary}\label{cor:cc_tractable}
For each pair of matrix functions $(W_\tau(x_{\tau,k}),L_\tau(x_{\tau,k}))$ satisfying instances of the condition
\begin{equation}\label{eq:tractable_con}
 \begin{bmatrix}
            W_{\tau,+} & A_\tau W_\tau +B_\tau L_\tau \\
           (A_\tau W_\tau +B_\tau L_\tau)^\top  & (1-\beta_\tau) W_\tau
        \end{bmatrix} > 0,
\end{equation}
where $W_\tau = M_\tau^{-1}$, $L_\tau=K_\tau W_\tau$, $W_{\tau,+}=W_\tau(x_{\tau,k+1})$, $M_\tau$ is uniformly bounded as $\underline{m}_\tau I \leq M_\tau(x_{\tau,k}) \leq \overline{m} _\tau I$ and $\beta_\tau \in (0,1]$, $M_\tau$ is then a discrete-time-control contraction metric and $K_\tau$ is a contracting differential feedback gain for \eqref{eq:sys_nom_taus}--\eqref{eq:dsys_fed_taus} over intervals of $\tau$.
\end{corollary}
\begin{proof}
Define $W_\tau(x_{\tau,k}) := M_\tau^{-1}(x_{\tau,k})$, $W_\tau(x_{\tau,k+1}) := M_\tau^{-1}(x_{\tau,k+1})$, $L_\tau(x_{\tau,k}) := K_\tau(x_{\tau,k})W_\tau(x_{\tau,k})$, apply Schur's complement to \eqref{eq:dist ccon}, and left/right multiply by $\diag\{I,W_\tau\}$ (and its transpose) to obtain \eqref{eq:tractable_con}. 
\end{proof}
\begin{remark}
Offline solution to \eqref{eq:tractable_con} can be  completed, e.g., via the methods in \cite{wei2021contractionsyntehsis,wei2021discrete}, which in turn implies problem feasibility (in the sense of stabilising controller existence), whereby one particular feasible controller (unconstrained) is given by the feedback control structure obtained by integrating \eqref{eq:dsys_fed_taus} along the geodesic $\gamma_\tau(x_{\tau,k},x^*_{\tau,k})$, i.e.,
\begin{equation}\label{equ:ctl int taus}
    u_{\tau,k} = u^*_{\tau,k} + \int_0^1 K_\tau(\gamma_\tau(s))\delta_{\gamma_{\tau,k}(s)} \,\,ds.
\end{equation}
To additionally ensure physical constraint satisfaction, a common approach is to search offline for less aggressive contraction rates, resulting in smaller control action magnitudes, and hence comply with the system constraints (see, e.g., \cite{mccloy2021differential}).
\end{remark}

%%%%%%%%%%%%%%%%%%%%%%%%%%%%%%%%%%%%%%%%%%%%%%%%%%%%%%%%%%%%%%%%%%%%%%%%%%%%%%%%%%%%%%%%%%%%%%%%%%%%%%%%%%%%
\subsection{Contraction-based Stability Constraint}
%\section{Contraction-constrained MPC with Disturbance Forecasting}
Optimal solutions to the MPC problem \eqref{eq:EMPCini} can be associated with state values which deviate from target operating points. As a consequence, the MPC requires a contraction based constraint or equivalent to ensure reference tracking, as will be constructed in the following.
\begin{corollary}\label{cor:cc_predH}
For \eqref{eq:sys_nom_taus}, the contraction constraint 
\begin{equation}\label{eq:shrinking_Rdist}
\begin{split}
&r(x_{\tau,k},x^*_{\tau,k},u_{\tau,k},\beta_\tau):= d\left(\gamma_\tau(x_{\tau,k+1},x^*_{\tau,k+1})\right) \\&\qquad - \left(1-\beta_\tau) \right)^{1/2} d\left(\gamma_\tau(x_{\tau,k},x^*_{\tau,k})\right) \leq 0, 
\end{split}
\end{equation}
where the geodesic $\gamma_\tau$ is computed using \eqref{equ:geodesic} with respect to the timescale specific metric $M_\tau$, defines the set of contracting tracking controllers which ensure the system state, $x_{\tau,k}$, is exponentially convergent to the feasible trajectory, $x^*_{\tau,k}$. 
\end{corollary}
\begin{proof}
Integrating \eqref{eqe:Vdotcondition} with \eqref{equ:generalised distance} along the geodesic $\gamma_\tau(x_{\tau,k}, x_{\tau,k}^*)$ yields (with a slight abuse of notation)
\begin{equation}\label{equ:energy bound2}
\begin{aligned}
        &\int_0^1 \delta_{\gamma_{\tau,k+1}(s)}^\top M_\tau(\gamma_{\tau,k+1}(s))\delta_{\gamma_{\tau,k+1}(s)} \,\,ds \\&\qquad\leq \int_0^1 (1-\beta_\tau) \delta_{\gamma_{\tau,k}(s)}^\top M_\tau(\gamma_{\tau,k}(s))\delta_{\gamma_{\tau,k}(s)} \,\,ds.
\end{aligned}
\end{equation}
Using \eqref{equ:Riemannian distance and energy}, \eqref{equ:geodesic}, \eqref{equ:energy bound2}, the property $d(\gamma)^2 \leq d(c)^2$, and taking square roots gives the condition in \eqref{eq:shrinking_Rdist}. 
\end{proof}
\begin{remark}
The contraction constraint in \eqref{eq:shrinking_Rdist}, has a reference-independent structure and is a nonlinear function affine in the control action, whereby the information required for the computation of $d(\gamma_\tau)$ is available from the current state measurement and knowledge of the system model and reference dynamics. In addition, these conditions define a set (or the conditional property) of controllers which guarantee the system is contracting (under Corollary \ref{cor:cc_taus}), whereby \eqref{equ:ctl int taus} provides one particular solution. 
\end{remark}

%%%%%%%%%%%%%%%%%%%%%%%%%%%%%%%%%%%%%%%%%%%%%%%%%%%%%%%%%%%%%%%%%%%%%%%%%%%%%%%%%%%%%%%%%%%%%%%%%%%%%%%%%%%%
\subsection{Constructing the MPC}
The multistage optimisation cost in \eqref{eq:EMPCini} can be reparameterised using the more convenient form in \eqref{eq:sys_nom_taus} (cf. \eqref{eq:sys_dist_nom}) as
\begin{equation}\label{eq:multi_cost}
     \sum_{i=0}^{N} \ell (\hat{x}_i,\hat{u}_i,\tau_i) = \sum^p_{i=1}  \sum^{N_{i}}_{j = 0} \ell(\hat{x}_{i,j},\hat{u}_{i,j},\tau_i),
\end{equation}
whereby the prediction horizon of length $N$ has been broken into (non necessarily uniform) segments based on the desired number of predictive steps using each time interval, i.e., 
\begin{equation}\label{eq:tausteps}
    N = \sum_{i=1}^p N_i, \qquad \hat{k} = \sum_{i=1}^p N_i \cdot \frac{\tau_i}{\tau_\Delta},
\end{equation}
whereby $\hat{k}$ denotes the total number of predicted discrete time-steps forward for the actual system in \eqref{eq:sys_dist_nom}. For example, given that the control actions for \eqref{eq:sys_dist_nom} are implemented on intervals of $\tau_\Delta$, suppose that $p=2$, $\tau_1 = \tau_\Delta$, $\tau_2 = 5\cdot\tau_\Delta$ and $N_1=N_2=1$. Then, in this scenario, the total number of predicted discrete time-steps forward for \eqref{eq:sys_dist_nom} would be $\hat{k} = 6$. Relative to a uniform timescale model ($\hat{k} = N = 2$) this approach offers an increased number of prediction steps without introducing additional complexity (or, by choosing an equivalent $\hat{k}$, at reduced computational complexity). 

Each subsequent predictive model (corresponding to a specific $\tau_i$) used in \eqref{eq:multi_cost} is initialised with the last predicted state of the previous model, i.e., 
\begin{equation}\label{eq:multi_stage_state}
    \hat{x}_{1,0} = x_k, \qquad \hat{x}_{i,0} = \hat{x}_{i-1,N_j}, \quad i=2,\cdots,p
\end{equation}
and hence each sequence of predicted control inputs is defined as the concatenation along each segment, i.e., 
\begin{equation}\label{eq:multi_stage_control}
\begin{split}
\mathbf{\hat{u}} &= (\hat{u}_0,\cdots,\hat{u}_{N-1}) \\ &=\left((\hat{u}_{1,0},\dots,\hat{u}_{1,N_1-1}),(\hat{u}_{2,0},\dots,\hat{u}_{2,N_2-1}),\right.\\&\left.\qquad\cdots,(\hat{u}_{p,0},\dots,\hat{u}_{p,N_p-1}) \right)
\end{split}
\end{equation}

From \eqref{eq:sys_nom_taus}, \eqref{eq:shrinking_Rdist}, \eqref{eq:multi_cost}--\eqref{eq:multi_stage_control} we can construct the contraction-constrained multi-timescale MPC optimisation problem:
\begin{equation}
\begin{aligned}\label{eq:EMPC}
\quad \min_{\mathbf{\hat{u}}} & \sum^p_{i=1}  \sum^{N_{i}}_{j = 0} \ell(\hat{x}_{i,j},\hat{u}_{i,j},\tau_i), \\
\text{s.t.} \quad & \hat{x}_{1,0} = x_k, \quad \hat{x}_{i,0} = \hat{x}_{i-1,N_j},  \\
\quad & \hat{x}_{i,j+1} = f_i(\hat{x}_{i,j}) + B_i\hat{u}_{i,j}, \\
& \hat{u}_i \in \mathcal{U}, \quad \hat{x}_i \in \mathcal{X}, \\ &r(\hat{x}_{i,j},x^*_{i,j},\hat{u}_{i,j},\beta_{\tau_i}) \leq 0
\end{aligned}
\end{equation}

For the MPC in \eqref{eq:EMPC}, we then have the following. 
\begin{theorem}\label{thm:mainmpc}
    An MPC with solution to \eqref{eq:EMPC} achieves economic optimisation simultaneously with reference tracking for multiple timescale nonlinear systems of the form in \eqref{eq:sys_dist_nom}. 
\end{theorem}
\begin{proof}
By considering a discrete number of time intervals, $\tau_i$, $i=1,\cdots,p$, \eqref{eq:sys_dist_nom} can be described (potentially exactly) by $p$ models of the form in \eqref{eq:sys_nom_taus}. For each $i$-th model, the pair $(M_{\tau_i},K_{\tau_i})$ are found satisfying \eqref{eq:dist ccon} (e.g., via solution to \eqref{eq:tractable_con}). 
Reference convergence over each horizon $N_i$ is ensured since economic cost minimising solutions to \eqref{eq:EMPC} are constrained to the set of control actions which possess the contracting property under Corollary \ref{cor:cc_predH}. Since each subsequent predictive model (used over the horizon of length $N_i$) is initialised with the last predicted state of the previous model via \eqref{eq:multi_stage_state}, the resulting sequence of predicted states (governed by a predicted control sequence of the form in \eqref{eq:multi_stage_control}) reparameterised from \eqref{eq:sys_nom_taus} and concatenated along each segment
\begin{equation}\label{eq:multi_stage_predicted_state}
\begin{split}
\mathbf{\hat{x}} &=\left((\hat{x}_{1,0},\dots,\hat{x}_{1,N_1}),(\hat{x}_{2,0},\dots,\hat{x}_{2,N_2}),\right.\\&\left.\qquad\cdots,(\hat{x}_{p,0},\dots,\hat{x}_{p,N_p}) \right)\\
&= (\hat{x}_0,\cdots,\hat{x}_{N})
\end{split}
\end{equation}
converges to the concatenated reference sequence reparameterised from \eqref{eq:sys_ref_taus}
\begin{equation}\label{eq:multi_stage_reference}
\begin{split}
\mathbf{x}^* &= (x^*_0,\cdots,x^*_{N})\\
&=\left((x^*_{1,0},\dots,x^*_{1,N_1}),(x^*_{2,0},\dots,x^*_{2,N_2}),\right.\\&\left.\qquad\cdots,(x^*_{p,0},\dots,x^*_{p,N_p}) \right),
\end{split}
\end{equation}
i.e., $d(\hat{x},x^*) \to 0$ and hence $\hat{x} \to x^*$. Initialising the predicted state sequence using the current state measurement, i.e., $\hat{x}_{1,0} = x_k$, and implementing the controller in receding horizon fashion \eqref{eq:EMPCu}, ensures $x_k \to x^*_k$ in the absence of prediction errors.  
\end{proof}
\begin{remark}
Prediction accuracy depends on choosing an adequate number of timescale specific models (many processes are naturally described by an infinite number of timescales). For a finite number of intervals and hence models, the effects of prediction errors on the reference tracking error can be attenuated by using the differential dissipativity techniques developed in \cite{wei2021contractionsyntehsis} and imposing additional requirements during metric synthesis (via a truncated incremental $\mathcal{L}_2$ gain). 
\end{remark}

% \begin{remark}
% % \RM{discuss that the geodesic associated with each metric at the crossover point doesn't need to be equal, we just need contraction on the corresponding horizon for each metric etc.}
% \end{remark}

%%%%%%%%%%%%%%%%%%%%%%%%%%%%%%%%%%%%%%%%%%%%%%%%%%%%%%%%%%%%%%%%%%%%%%%%%%%%%%%%%%%%%%%%%%%%%%%%%%%%%%%%%%%%
\subsection{Discussion}

The proposed MPC approach achieves economic optimisation with stability for multiple timescale nonlinear systems. In particular, this carries several advantages:
\begin{itemize}
    \item Economic reference tracking -- the optimisation problem permits arbitrary economic cost functions, with reference tracking ensured by contraction constraints. 
    \item Reference flexibility -- the contraction (stability) constraint is structurally reference-independent, constructed offline and updated online as new state measurements and reference information is available.
    \item Reduced computational burden -- with respect to an equivalent single timescale approach, longer prediction horizons can be considered by using longer time intervals (and incorporating appropriate slower dynamic models). 
    \item Scalability via \textit{modular} stability constraint construction -- contraction constraints can be systematically added to the optimisation problem as additional time intervals (and hence models) are considered, without increasing offline synthesis or online computational requirements for an equivalent prediction length (each constraint is formed and imposed independently).
\end{itemize}

In future work, natural extensions can be made to the existing framework to simultaneously incorporate reduced order models (e.g., by adjusting model granularity \cite{findeisen2016exploiting}). Feasibility analysis, extension to a more general class of nonlinear systems and explicit consideration for prediction / timescale discretisation errors are also part of future studies.

%%%%%%%%%%%%%%%%%%%%%%%%%%%%%%%%%%%%%%%%%%%%%%%%%%%%%%%%%%%%%%%%%%%%%%%%%%%%%%%%%%%%%%%%%%%%%%%%%%%%%%%%%%%%
\section{Illustrative Example}\label{sec:exa}
In the following numerical example, we consider the coupled-tank process described in~\cite{mccloy2014set}, consisting of two cylindrical tanks positioned one directly above the other. Liquid flows from an upper primary tank through a lower pilot tank to a reservoir, and a pump is used to push liquid from the reservoir back up to the upper tank. The objective is to minimise the pump energy input whilst achieving a target liquid level in the primary tank (matched by the accessible secondary tank). The discrete-time dynamics of the liquid levels $x_1$ (upper tank) and $x_2$ (lower tank) are described as
\begin{equation} \label{eq:coupledTanks}
\begin{aligned}
x_{1,k+1} & = x_{1,k} -\tau\frac{\sigma_1}{\alpha_1} \sqrt{2g x_{1,k}} + \tau\frac{k_p}{\alpha_1} u_k,\\
x_{2,k+1} & = x_{2,k} +\tau\frac{\sigma_2}{\alpha_2} \sqrt{2g x_{1,k}} - \tau\frac{\sigma_2}{\alpha_2}
\sqrt{2g x_{2,k}},
\end{aligned}
\end{equation}
where $u_k$ is the voltage applied to the pump, and the parameters are as follows:
$\alpha_1=155$cm$^2$ and $\alpha_2=15.5$cm$^2$ are the cross-sectional areas of tanks 1 and 2; $\sigma_1=\sigma_2= 0.178$cm$^2$ are the cross-sectional areas of each tank's outflow
orifice; $k_p=13.2$cm$^3$/Vs is the gain of the pump; $g=980$cm/s$^2$; and $\tau_\Delta = 1s$ is the natural sampling period of the system. The liquid levels are constrained to the range $x_1,x_2 \in [0,10]$cm, and the pump voltage is limited to $u \in [0,10]$V. 

To capture the relatively fast and slow dynamics of each tank, two timescales ($p=2$ models), $\tau_1 = 1s$ and $\tau_2 = 10s$, are considered. Control contraction metrics are then synthesised via the following Sum-of-Squares programming formulation (see \cite{wei2021control} for more detail), 
\begin{equation}\label{min:sos}
    \begin{aligned}
        &\underset{l_{\tau,c},w_{\tau,c},r}{\min} r\\
        &s.t. \ \phi^\top  \Omega_\tau \phi -rI \in \Sigma(x_k,u_k,\phi), ~r \geq 0,
    \end{aligned}
\end{equation} 
where $\Omega_\tau$ is formed by the left hand side of \eqref{eq:tractable_con}, $\Sigma(x_k,u_k,\phi)$ is a Sum-of-squares function of the arguments $x_k,u_k$ and $\phi$, $\phi$ is a vector of monomials, $l_{\tau,c},w_{\tau,c}$ are the polynomial coefficients of elements of $W_\tau$ and $L_\tau$ in \eqref{eq:tractable_con}, and $r$ is a small positive constant. Solution pairs $(W_1,L_1)$ and $(W_2,L_2)$ to the Sum-of-Squares programming problem in \eqref{min:sos} satisfy \eqref{eq:tractable_con} and consequently the contraction condition in \eqref{eq:dist ccon} using the definitions $W=M^{-1},L=KW$. Hence, the pairs $(M_1,K_1)$ and $(M_2,K_2)$ were obtained from solutions to \eqref{min:sos}, for the two respective time scale systems, $\tau = 1$ and $\tau = 10$, with contraction rates $\beta_1=\beta_2 = 0.3$. 

The MPC is then formulated as in \eqref{eq:EMPC} with two contraction metrics $M_1,M_2$ and two prediction horizons $N_1 = 1$, $N_2 = 3$ (corresponding to the two time intervals $\tau_1$ and $\tau_2$). The resulting total number of predicted time steps was computed via \eqref{eq:tausteps} as $\hat{k} = N_1 \cdot \frac{\tau_1}{\tau_\Delta} + N_2 \cdot \frac{\tau_2}{\tau_\Delta} = 31$. An economic cost function was designed to reflect the control effort, i.e., 
%\begin{equation}
$\ell = ||u_k||^2$, 
%\end{equation}
subject to the time-varying operating targets: $(x_1^*,x_2^*)$=$(2.5,2.5)$, $(5,5)$ and $(7.5,7.5)$ on the respective intervals of  $[0,40)$, $[40,80)$, $[80,120]s$. Constraint feasibility of \eqref{eq:shrinking_Rdist}, under the contraction rates $\beta_1,\beta_2$, was verified via constraint sampling of $u \in \mathcal{U}$ using $(M_1,K_1)$, $(M_2,K_2)$, the controller \eqref{equ:ctl int taus} and model \eqref{eq:coupledTanks} for $x\in \mathcal{X}$. 

The closed-loop system was then simulated with the MPC \eqref{eq:EMPC} computed for each $\tau_\Delta=1s$ interval. The average computation time, per simulated step, for the proposed MPC, 'MPC($\tau_1,\tau_2$)', is compared with an equivalent single timescale MPC, 'MPC($\tau_\Delta$)', which, as shown in Table \ref{tbl:compt}, indicates that the proposed method requires significantly less computational effort for an equivalent number of prediction steps on average (in fact, the single timescale method is infeasible under the specified control period of $1s$).
\begin{table}[h]\caption{Computation Time Comparison.}\label{tbl:compt}
\centering
\begin{tabular}{c|c|c}
    Method & Predicted Steps & Ave. Comp. Time \\
    \hline
    MPC($\tau_1,\tau_2$) & $\hat{k} = 31$ & $0.425s$\\
    MPC($\tau_\Delta$) & $\hat{k} = 31$ & $3.106s$\\
\end{tabular}
\end{table}

\begin{figure}\label{fig:sim}
\centering
        \includegraphics[width=\linewidth]{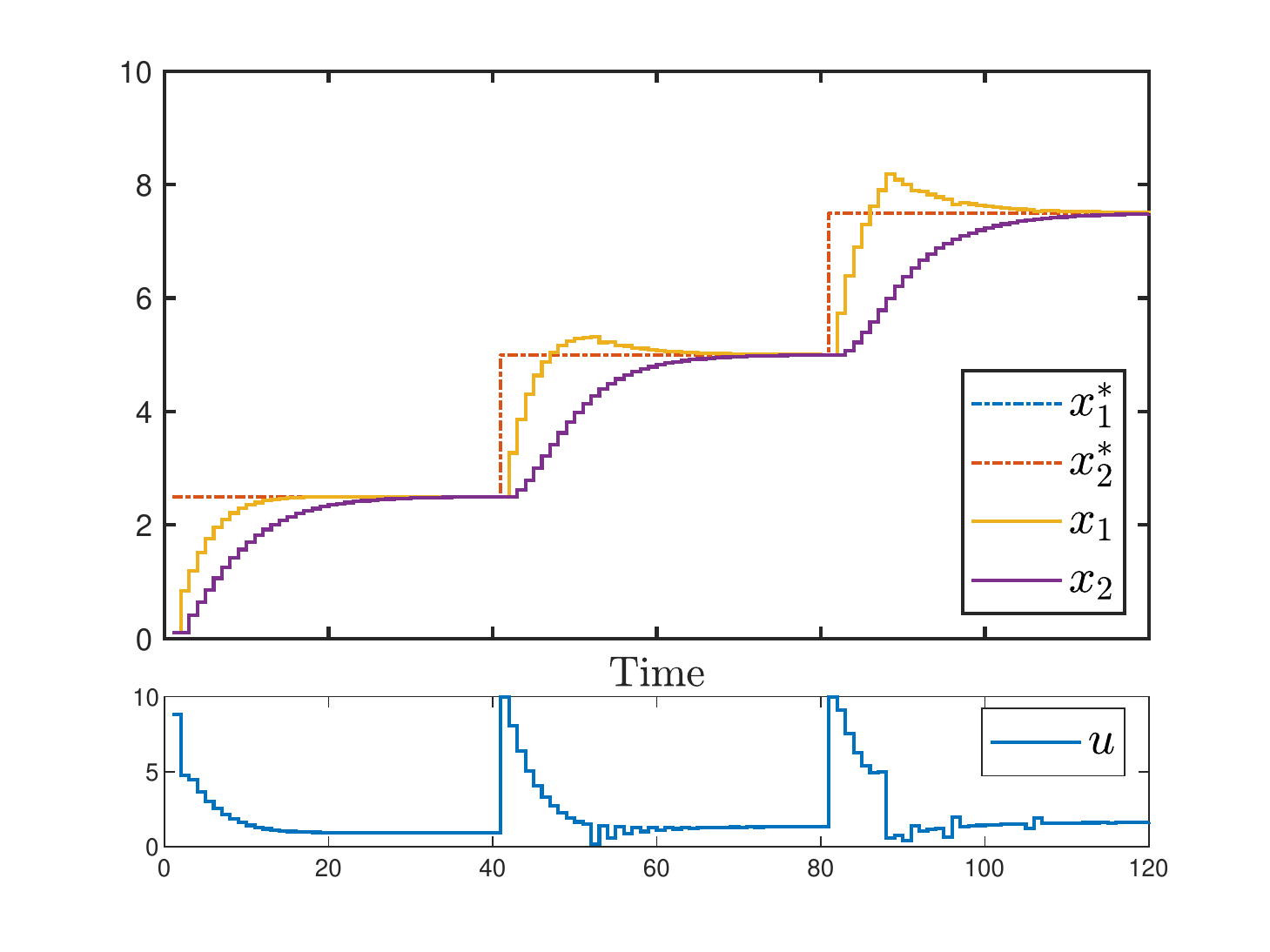}
        \caption{Simulation results for control of \eqref{eq:coupledTanks} using \eqref{eq:EMPC}.}
            \squeezeup
\end{figure}

Finally, the simulation results shown in Fig. \ref{fig:sim}, illustrate that the proposed contraction constrained multi-timescale MPC is capable of driving the system to the operational targets whilst minimising the input energy as per Theorem \ref{thm:mainmpc}.   

%%%%%%%%%%%%%%%%%%%%%%%%%%%%%%%%%%%%%%%%%%%%%%%%%%%%%%%%%%%%%%%%%%%%%%%%%%%%%%%%%%%%%%%%%%%%%%%%%%%%%%%%%%%%
\section{Conclusion} \label{sec:conclusion}
A contraction constrained MPC approach was developed for discrete-time nonlinear processes with multiple timescale dynamics. Through the differential system framework, stability conditions were derived to ensure convergence of the resulting closed-loop to feasible time-varying references. These conditions were also generalised to characterise a set of stabilising controllers and imposed as stability constraints on an MPC. The MPC would then search for a cost minimising controller amongst those stabilising controllers whilst utilising multiple timescale models. The result was an MPC capable of ensuring convergency of multiple timescale nonlinear processes. The overall approach was effectively demonstrated via simulation.

\bibliographystyle{IEEEtran}
\bibliography{CDC22_main.bib}

\end{document}